\documentstyle[prl,aps,floats]{revtex}  

\begin{document}  
\preprint{astro-ph/0006421} 
\draft 
  
\input epsf 
  
\renewcommand{\topfraction}{0.99} 
\renewcommand{\bottomfraction}{0.99} 
  
\twocolumn[\hsize\textwidth\columnwidth\hsize\csname 
@twocolumnfalse\endcsname 
  
\title{Steep inflation: ending braneworld inflation by gravitational particle 
production} 
\author{Edmund J. Copeland$^1$, Andrew R.~Liddle$^2$ and James E.~Lidsey$^3$}
\address{$^1$Centre for Theoretical Physics, CPES, University of Sussex, 
Brighton BN1 9QJ, United Kingdom} 
\address{$^2$Astronomy Centre, University of Sussex, Brighton BN1 9QJ, United 
Kingdom} 
\address{$^3$Astronomy Unit, School of Mathematical Sciences, Queen Mary and 
Westfield, Mile End Road, London E1 4NS, United Kingdom}
\date{\today} 
\maketitle 
\begin{abstract}  
We propose a scenario for inflation based upon the braneworld picture, in which 
high-energy corrections to the Friedmann equation permit inflation to take place 
with potentials ordinarily too steep to sustain it. Inflation ends when the 
braneworld corrections begin to lose their dominance. Reheating may naturally be 
brought about via gravitational particle production, rather than the usual 
inflaton decay mechanism; the reheat temperature may be low enough to 
satisfy the gravitino bound and the Universe becomes radiation dominated early 
enough for nucleosynthesis. We illustrate the idea by considering steep 
exponential potentials, and show they can give satisfactory density 
perturbations (both amplitude and slope) and reheat successfully. The scalar 
field may survive to the present epoch without violating observational bounds, 
and could be invoked in the quintessential inflation scenario of Peebles and 
Vilenkin.
\end{abstract}   
  
\pacs{PACS numbers: 98.80.Cq \hfill astro-ph/0006421} 
  
\vskip2pc] 
  
\section{Introduction}  

The realization \cite{ark} that we may live on a so-called brane embedded in a 
higher-dimensional Universe has significant implications for cosmology. One such 
is that the Friedmann equation is modified at very high energies 
\cite{bin,shiro}, acquiring a term quadratic in the density. Such a term 
generally makes it easier to obtain inflation in the early Universe, by 
contributing extra friction to the scalar field equation of motion 
\cite{maart,quadref}. 

In this {\it Letter} we propose to capitalize on this feature by considering
potentials which are normally too steep to support inflation.  As we will see,
this allows a particularly natural implementation of an unusual model for
reheating at the end of inflation.  In the conventional inflationary scenario,
reheating is achieved by decay of the inflaton field into normal matter, either
via parametric resonance or by single particle decays.  However, there are at
least two other mechanisms which have been proposed in the literature.  One is
reheating via the production of primordial black holes \cite{pbh}, whose
subsequent Hawking evaporation produces the conventional matter --- we do not
discuss this possibility further.  The second, which we exploit here, is
that particles arise from gravitational particle production, with the inflaton
energy density subsequently redshifting sufficiently quickly that this produced
radiation comes to dominate.  This scenario was first discussed by Ford
\cite{Ford} (see also Ref.~\cite{GS}) and refined by Spokoiny \cite{Spok};
recent applications include baryogenesis \cite{JP}, scaling solutions in the
present Universe \cite{FJ}, and the `quintessential inflation' scenario of
Peebles and Vilenkin \cite{PV,FKL}.

In conventional inflationary models, the gravitational production scenario is
difficult to arrange.  In order to redshift the inflaton density more quickly
than that of the produced particles one needs a sharp feature in the potential,
so that it becomes steep enough that the inflaton becomes kinetic energy
dominated, leading to $\rho_\phi \propto 1/a^6$.  The gravitational particle
production scenario arises much more naturally in the context of inflation on
the brane.  When one considers a potential which would ordinarily be too steep
to support inflation, the transition to a kinetic energy dominated regime occurs
naturally once the energy density falls sufficiently for the quadratic term to
begin to lose its dominance.  Thus there is no need for a feature in the
inflaton potential.


Although our scenario could be implemented for a wide range of inflaton 
potentials, for simplicity we focus our discussion on the well-studied 
exponential potentials \cite{LM}
\begin{equation}
V(\phi) = V_0 \exp \left( - \sqrt{8\pi} \, \alpha \frac{\phi}{M_4} \right) \,,
\end{equation}
where $M_4$ is the four-dimensional Planck mass and $\alpha$ is a constant. In 
the standard cosmology, the potential is shallow enough to support inflation 
only if $\alpha^2 < 2$.

We consider the five-dimensional brane scenario, in which the Friedmann equation 
is modified from its usual form, becoming \cite{bin,shiro}
\begin{equation}
H^2 = \frac{8\pi}{3 M_4^2} \, \rho \, \left[ 1 + \frac{\rho}{2\lambda} \right]
	+ \frac{\Lambda_4}{3} + \frac{{\cal E}}{a^4} \,,
\end{equation}
where $\Lambda_4$ is the 
four-dimensional cosmological constant and the final term 
represents the influence of bulk gravitons on the brane. The brane tension 
$\lambda$ relates the four and five-dimensional Planck masses via
\begin{equation}
M_4 = \sqrt{\frac{3}{4\pi}} \, \left( \frac{M_5^2}{\sqrt{\lambda}} \right) \,
	M_5 \,,
\end{equation}
and is constrained by the requirement of successful nucleosynthesis as $\lambda 
> (1 \, {\rm MeV})^4$. We assume that the 
four-dimensional cosmological constant is set to zero by some (as yet 
undiscovered) mechanism, and once inflation begins the final term will rapidly 
become unimportant, leaving us with
\begin{equation}
H^2 = \frac{8\pi}{3 M_4^2} \, \rho \, \left[ 1 + \frac{\rho}{2\lambda} 
	\right] \,.
\end{equation}
We assume that the scalar field is confined to the brane, so that its field 
equation has the standard form
\begin{equation}
\ddot{\phi} + 3 H \dot{\phi} = - \frac{dV}{d\phi} \,.
\end{equation}

\section{Constraints}

We analyze our model following closely the formalism of Maartens et 
al.~\cite{maart}. As we will see later, suitable values of $\alpha$ will be 
greater than about five; such potentials would not give inflation in the 
conventional cosmology.

\subsection{Inflationary dynamics}

Following Maartens et al.~\cite{maart}, we define a slow-roll parameter 
$\epsilon$, generalizing the usual ones 
\cite{LL}, by
\begin{equation}
\epsilon \equiv - \frac{\dot{H}}{H^2} \simeq \frac{M_4^2}{4 \pi} \, 
	\left( \frac{V'}{V} \right)^2 \,
	\frac{1+V/\lambda}{\left(2 + V/\lambda \right)^2} \,,
\end{equation}
where prime indicates a $\phi$-derivative and the slow-roll approximation has 
been employed.
Inflation takes place whenever $\epsilon < 1$. For our model we can always take 
$V \gg \lambda$ during inflation, so
\begin{equation}
\epsilon \simeq \frac{2\alpha^2\lambda}{V} \,.
\end{equation}
The end of inflation will take place when $\epsilon = 1$, giving
\begin{equation}
\label{Vend}
V_{{\rm end}} \simeq 2\alpha^2 \lambda \,.
\end{equation}
Notice that even at the end of inflation, typically the term quadratic in the 
density still dominates the linear term by a significant factor. 

The amount of inflation that takes place is given by the number of $e$-foldings 
\cite{maart}
\begin{equation}
N \simeq - \frac{8\pi}{M_4^2} \int_\phi^{\phi_{{\rm end}}} \frac{V}{V'} \, 
\left(1 + 
\frac{V}{2\lambda} \right) \, d\phi \,.
\end{equation}
As $V \gg \lambda$ during inflation, this simply becomes
\begin{equation}
N \simeq \frac{1}{2\lambda \alpha^2} \left(V_N - V_{{\rm end}} \right) \,,
\end{equation}
and so the potential $V_N$ a given number of $e$-foldings from the end of 
inflation is given by
\begin{equation}
\label{VN}
V_N = V_{{\rm end}} \left( N+1 \right) \,.
\end{equation}

\subsection{Perturbations and the COBE normalization}

The amplitude of density perturbations is given by \cite{maart}
\begin{equation}
A_{{\rm S}}^2 \simeq \frac{512 \pi}{75 M_4^6} \, \frac{V^3}{V'^2} \, 
	\left( 1+\frac{V}{2\lambda} \right)^3 \simeq \frac{8}{75} 
	\, \frac{V^4}{M_4^4 \alpha^2 \lambda^3} \,.
\end{equation}
The observed value from COBE is $A_{{\rm S}} = 2 \times 10^{-5}$ 
\cite{COBEnorm}, which we apply 
50 $e$-foldings from the end of inflation; using Eqs.~(\ref{Vend}) and 
(\ref{VN}) we obtain the simple result
\begin{equation}
\label{COBE1}
V_{{\rm end}}^{1/4} = \frac{9 \times 10^{-5}}{\alpha} \, M_4 = \frac{1 \times 
10^{15} \, {\rm GeV}}{\alpha} \,.
\end{equation}
Equivalently, this fixes the brane tension as
\begin{equation}
\label{COBE2}
\lambda \simeq \frac{4 \times 10^{-17}}{\alpha^6} \, M_4^4 =
	\left(\frac{10^{15} \, {\rm GeV}}{\alpha^{3/2}}\right)^4 \,,
\end{equation}
or the five-dimensional Planck mass as 
\begin{equation}
\label{COBE3}
M_5 = \frac{2.3 \times 10^{-3}}{\alpha} \, M_4 = \frac{3 \times 10^{16} 
	\, {\rm GeV}}{\alpha} \,.
\end{equation}

For consistency it should be verified that enough inflation can
indeed occur.  The assumption that the scalar field is confined to the brane
becomes unreliable if $V > M_5^4$.  Imposing this condition along with the COBE
normalization leads to the simple result 
\begin{equation} 
N_{{\rm max}} = 4 \times 10^5 \,, 
\end{equation} 
independent of both $\alpha$ and $\lambda$.  This
large dimensionless number has its origin in the small dimensionless number
corresponding to the observed temperature anisotropy.  It may be that more 
inflation is
possible before the scalar field becomes confined to the brane, but the 
observable last 50 or so $e$-foldings will be in that regime.

The spectral index of the density perturbation spectrum can also be found 
following Ref.~\cite{maart}, leading to
\begin{equation}
n = 1 - \frac{4}{N+1} \,.
\end{equation}
This result is independent of both $\alpha$ and the brane tension, and with the 
usual assumption of 50 $e$-foldings gives $n = 0.92$. This is compatible with 
current observations (e.g.~see Ref.~\cite{Boom}), but can be strongly tested in 
future.

In the high-energy limit of the braneworld scenario the contribution of 
gravitational waves relative to density perturbations is suppressed, and 
the relative amplitude is given by \cite{LMW}
\begin{equation}
\frac{A_{{\rm T}}^2}{A_{{\rm S}}^2} \simeq \frac{3 M^2_4}{16\pi} \,
	\left(\frac{V'}{V} \right)^2 \,	\frac{2\lambda}{V} \simeq 0.03 \,,
\end{equation}
which again is independent of all model parameters. This is a significant level 
of production, not far from current upper limits; in terms of the commonly-used 
quantity $r \simeq 4\pi A_{{\rm T}}^2/A_{{\rm S}}^2$ we have $r \simeq 0.4$, and 
it is believed that the Planck satellite will eventually measure $r$ with an 
error bar of around 0.05. 

\subsection{Reheating by gravitational particle production}

At the end of inflation, the scalar field density is given by $V_{{\rm end}}$. 
Some of this energy needs to be converted to conventional matter to restore the 
hot big bang cosmology. Usually this is brought about by decay of the inflaton 
field, though this could only proceed to completion if the potential is modified 
to introduce a minimum.

Instead, we take advantage of the possibility of reheating by gravitational 
particle production \cite{Ford,Spok,FJ}, where the required particles are 
produced 
quantum mechanically from the time-varying gravitational field. Standard 
calculations \cite{Ford,Spok} give the density of particles produced at the end 
of inflation as
\begin{equation}
\rho_{{\rm R}} \sim 0.01 g_{{\rm prod}} H_{{\rm end}}^4 \simeq  0.2  \,
	g_{{\rm prod}} \, \frac{V_{{\rm end}}^4}{\lambda^2 M_4^4} \,,
\end{equation}
where $g_{{\rm prod}}$ is the number of fields experiencing particle production, 
likely to be 
between 10 and 100. This result depends only on the fields' equations of motion 
and the expansion rate, and so remains valid in the braneworld scenario. The 
relative densities at the end of inflation are
\begin{equation}
\label{initrat}
\frac{\rho_{{\rm R}}}{\rho_\phi} = \, 0.2 g_{{\rm prod}} \frac{V_{{\rm
	end}}^3}{\lambda^2 M_4^4} = 5 g_{{\rm prod}} \times 10^{-17} \,,
\end{equation}
where the last equality uses the COBE normalization, Eqs.~(\ref{COBE1}) and 
(\ref{COBE2}). The numerical value of the radiation density, given the COBE 
normalization, is
\begin{equation}
\rho_{{\rm R}} \simeq g_{{\rm prod}} \left( \frac{10^{11} \, {\rm GeV}}{\alpha} 
	\right)^4 \,,
\end{equation}
which if immediately thermalized would give a temperature 
\begin{equation}
\label{reheat}
T_{{\rm end}} \simeq \frac{10^{11}\, {\rm GeV}}{\alpha} \, \left(
	\frac{g_{{\rm prod}}}{g_*} \right)^{1/4} \,,
\end{equation}
where $g_*$ is the total number of species in the thermal bath (which may be 
somewhat higher than $g_{{\rm prod}}$).
This temperature is around that at which thermal production of gravitinos via 
two-particle collisions becomes problematic \cite{Sarkar}, and only more 
detailed calculations (including the amount of cooling during the final 
thermalization) can clarify whether the gravitino limit can be satisfied. It 
also needs to be determined whether 
or not non-thermal gravitino production might be significant (see 
e.g.~Ref.~\cite{Lyth}). See Ref.~\cite{FKL} for other model-building 
difficulties that may arise in constructing a working scenario.

Although the ratio of radiation to scalar field 
densities at the end of 
inflation is small, it is possible for the radiation to become dominant. 
Once the high-energy correction to the Friedmann equation becomes 
unimportant, the potential is so steep that the evolution becomes completely 
kinetic-energy dominated, with $\rho_\phi \propto 1/a^6$. We first assume this 
behaviour sets in as soon as inflation ends, and examine the opposite case in 
the next paragraph. When the kinetic energy is completely dominant, we have
\begin{equation}
\label{Einrat}
\frac{\rho_{{\rm R}}}{\rho_\phi} \propto a^2 \,.
\end{equation}
{}From Eq.~(\ref{initrat}), we can estimate that the radiation comes to dominate 
after the 
Universe has expanded by a factor around $10^{7}$ to $10^8$ after inflation, at 
which stage the temperature, which obeys $T \propto 1/a$, is
\begin{equation}
\label{raddom}
T \simeq \frac{10^3 \, {\rm GeV}}{\alpha} \,.
\end{equation}
We see that radiation domination sets in comfortably before nucleosynthesis.

This needs some correction to allow for the period between the end of inflation
and the onset of validity of the usual Friedmann equation at $\rho \simeq 
2\lambda$,
which will cause a delay before Eq.~(\ref{Einrat}) applies.  As
inflation has ended, we know that $\rho_\phi$ must fall off at least as quickly
as $1/a$.\footnote{In the standard cosmology this would be $1/a^2$, coming from
the inflationary condition $p < -\rho/3$, but in the high-energy limit of brane
cosmology the inflationary condition becomes $p \lesssim -2 \rho/3$
\cite{maart}.}  This means a worst-case scenario where the scale factor has
increased by an extra factor of $(V_{{\rm end}}/2\lambda)^{5/6} = \alpha^{5/3}$
by the time the usual Friedmann equation sets in, as compared to the argument of
the previous paragraph.  The radiation temperature at that time is therefore
smaller by that factor, and so extra expansion by a factor $\alpha^{10/3}$, will
be required to bring on radiation domination.  For typical parameters this may
reduce the temperature at the onset of radiation domination by a factor between
a hundred and a thousand as compared to Eq.~(\ref{raddom}), which is still early
enough for successful nucleosynthesis. In any event, this worst-case scenario 
will
not be attained in practice, leading to a higher temperature at the onset of
radiation domination.

Notice that while in this cosmology the formal epoch of radiation domination 
does not begin until quite a low temperature [Eq.~(\ref{raddom}), corrected as 
described in the previous paragraph], the radiation 
originally was created at a much higher temperature [Eq.~(\ref{reheat})]. It may 
well 
be therefore that processes such as baryogenesis might occur within the 
radiation fluid at an epoch before it comes to dominate the density of the 
Universe --- see Ref.~\cite{JP}.

\subsection{Scaling in the present Universe}

The period of kinetic domination will not last indefinitely; it is well known 
that 
the late-time behaviour for steep exponential potentials is a scaling solution 
where 
the scalar field exhibits the same redshift dependence as the dominant fluid in 
the Universe \cite{scaling,CLW}. The scaling density, assuming spatial flatness, 
is
\begin{equation}
\Omega_\phi = \frac{3(w+1)}{\alpha^2} \,,
\end{equation}
where the dominant fluid has equation of state $p = w\rho$. If the scaling 
solution is already attained by nucleosynthesis, the field acts as extra 
relativistic 
species and will adversely affect the standard picture unless $\alpha^2 \gtrsim 
20$ \cite{FJ,CLW}. If the scaling solution is not attained by nucleosynthesis 
(as may 
well be the case as the kinetic energy has a tendency to overshoot the scaling 
solution by several orders of magnitude before approaching it from below 
\cite{FJ}) this 
limit can be evaded, but a similar limit can be derived from the effect 
of the scalar field on density perturbations at the present epoch \cite{FJ}.

Because the scalar field survives to the present, a suitable feature in the 
potential near its present value can give rise to quintessence behaviour; this 
is a version of the quintessential inflation scenario of Peebles and Vilenkin 
\cite{PV} in which the same scalar field gives both early Universe inflation and 
the present observed acceleration. More attractively, one could use a different 
form of the potential, such as an inverse power-law, which allows a natural 
transition to quintessence at late times and which merits further investigation. 
Our model allows such a scenario 
with a much simpler choice of potential than in Einstein gravity models due to 
our exploitation of the 
braneworld effects.

\section{Conclusions}
  
We have presented a non-standard model of inflation, made possible by 
braneworld-motivated corrections to the Friedmann equation at high energies. The 
novel ingredients are that inflation becomes possible for a class of 
potentials ordinarily too steep to sustain accelerated expansion, with the end 
of inflation brought 
on by the braneworld corrections losing their dominance. While standard 
reheating could be considered, this scenario allows a particularly natural 
implementation of reheating via gravitational particle production, and we have 
confirmed that all existing constraints can be satisfied.

For the particular implementation of this idea using steep exponential 
potentials, a distinctive prediction is that the spectral index of density 
perturbations is given 
by $n \simeq 0.92$, independent of all parameters (except the 
number of $e$-foldings to the end of inflation). This prediction, along with the 
large predicted amplitude of gravitational waves, is readily 
testable with upcoming experiments.
  
Although we have not attempted to place this scenario within the context of a
realistic particle physics model, it should be emphasized that current M-theory
models have a number of likely candidate scalar fields with steep exponential
potentials living on the brane.  These include the dilaton in five--dimensional
heterotic M-theory \cite{lukas98}, the dilaton present in self--tuning
mechanisms to cancel the cosmological constant on the brane \cite{horowitz2000},
and combinations of the dilaton and moduli fields that arise in heterotic
M-theory when studying the cosmological stabilization of these fields
\cite{barreiro2000}.  The important point is that all of these models would be
affected by the presence of the corrections to the Friedmann equation, and
intriguingly this could open up the possibility of inflation occurring before 
the
moduli fields become stabilized.

\section*{Acknowledgments}  
E.J.C.~was supported by PPARC, and J.E.L.~by the 
Royal Society and in part by the Sussex PPARC visitors 
grant. We thank Andre Lukas
and 
David Wands for useful discussions, and Lev Kofman and Alex 
Vilenkin for their comments.  
  
   
\end{document}